\documentstyle[preprint,prb,aps]{revtex}

\begin{document}
\draft
\title{A novel Microwave Frequency Scanning Capacitance Microscope }
\author{Atif Imtiaz,\thanks{%
email address: aimtiaz@squid.umd.edu}and Steven M. Anlage}
\address{Center for Superconductivity Research, Department of Physics, University of\\
Maryland, College Park, MD 20742-4111 }
\maketitle

\begin{abstract}
We report a new technique of scanning capacitance microscopy at microwave
frequencies. A near field scanning microwave microscope probe is kept at a
constant height of about 1 nm above the sample with the help of Scanning
Tunneling Microscope (STM) feedback. The microwaves are incident onto the
sample through a coaxial resonator that is terminated at one end with a
sharp tip (the same tip is used to conduct STM), and capacitively coupled to
a feedback circuit and microwave source at the other end. The feedback
circuit keeps the source locked onto the resonance frequency of the
resonator and outputs the frequency shift and quality factor change due to
property variations of the sample. The spatial resolution due to capacitance
variations is $\leq $ 2.5 nm. The microscope is broadband and experiments
were performed from 7 GHz to 11 GHz. We develop a quantitative transmission
line model that treats the tip to sample interaction as a series combination
of capacitance and effective sheet resistance in the sample.
\end{abstract}


{\bf Introduction:} Scanning Capacitance Microscopy can be used to map
spatial variations of the topography of conducting materials, or the
dielectric properites of thin films and bulk insulators\cite
{Matey1,Quate1,Williams}. In general, such microscopes detect a change in
capacitance, $\delta $C, by means of a resonant circuit that includes the
probe-sample capacitance. An early version of such a microscope employed a
diamond stylus in contact with the sample. When the stylus was touching the
sample, a probe electrode (attached to the stylus) was 20nm above the
sample. The electrode to sample capacitance was measured through the
changing resonant frequency of an inductor/capacitor (LC) resonant circuit%
\cite{Matey1}. With this microscope the lateral resolution claimed was 100nm
and vertical resolution of 0.3nm. Other capacitance microscopes have been
made by adding a similar resonant capacitance sensor to an Atomic Force
Microscope\cite{Quate1} and a Scanning Tunneling Microscope\cite{Williams}%
(STM). The lateral topographic resolutions reported were 75nm and 25nm,
respectively. These microscopes have been used for dopant profiling in
semiconductors\cite{Dransfeld,Williams2,Lowney}, for example.

However, in many materials of interest, measurement of loss is crucial to
extract the interesting physics. For example, certain colossal
magneto-resistance (CMR) materials have ''metallic'' and ''insulating''
phases that coexist on very small (almost 1-2 nm) length scales\cite{Dagotto}%
. In thin films of Bi$_2$Sr$_2$CaCu$_2$O$_{8+\delta }$, STM spectroscopic
data shows evidence for coexisting superconducting and semiconducting phases%
\cite{Klein2} on similar length scales. One way to distinguish between the
two phases is to locally measure ohmic losses. Existing capacitance
microscopes are not designed to image such quantities. A near-field
microwave microscope would be the right tool, and can be utilized to
quantitatively extract losses in the form of sheet resistance (R$_x$).
However, sub-micron spatial resolution is required to distinguish the finely
intermixed phases.

Our objective now is to obtain high-resolution loss imaging, in conjunction
with capacitance microscopy. The quantitative sheet resistance ($R_x$) of a
sample was extracted earlier\cite{Steinhauer4}, from the frequency shift and
quality factor of the microscope resonator with lateral resolution in the
10's to 100's of $\mu $m. We now want to improve the spatial resolution,
while maintaining the sensitive loss imaging capability.

With a scanning microwave microscope, one can illuminate a controlled
localized area with microwave fields and currents. In the past, some
near-field microwave microscopes have utilized STM\cite{Weiss,Keilmann} or
AFM tips\cite{vanderWeide,Cho,Cho2}simply to concentrate rf electric field
on a sample. The sharp end of the tip acts like a ''lightening rod'',
enhancing the spatial resolution for microwave microscopy. In an earlier
version of our microwave microscope, an STM tip was also used to focus
electric fields on the surface\cite{Steinhauer}. Because the tip was in
contact with the sample, with contact force of about 60 $\mu $N, the best
spatial resolution achieved was $\simeq $ 1 $\mu $m. The contact force
flattened the imaging end of the tip, increasing the radius of curvature.
One way to improve the spatial resolution of the microscope is to prevent
the tip from touching the sample. To achieve this, the new version of the
near-field microwave microscope (Fig. 1) has STM feedback integrated to
maintain a roughly 1 nm constant height during scanning.

{\bf Experiment: }The microscope, schematically presented in Fig. 1 is
similar to the versions discussed at length in prior publications\cite
{Steinhauer,Gus,Steinhauer5,Steinhauer2,Steinhauer3}. Changes made to our
microscope include using a bias tee to make the DC connection to STM
feedback while maintaining an AC coupling to the microwave source and
feedback circuit. Also, the sample is now on a XYZ\ piezoelectric (piezo)
translation stage, instead of an XY motor stage. At one end of the coaxial
resonator is an open ended coaxial probe with a sharp STM tip sticking out
of its center conductor. As with our previous microscopes\cite{Steinhauer},
the other end of the resonator is capacitively coupled to a microwave source
and a feedback circuit via a directional coupler and a diode detector, as
shown in Fig.1. The feedback circuit keeps the source locked onto the
resonance frequency $f_0$, of the resonator and it gives the frequency shift
($\Delta $f) and quality factor (Q) as output signals. However, the
probe-sample seperation is maintained by a constant current STM feedback
loop during scanning.

Many interesting materials have transition temperatures well below room
temperature, requiring a cryogenic microscope. We use a commercially
available Oxford Cryostat to cool the sample and part of the microscope. The
sample can be cooled to any temperature between 4.2K and room temperature.
The quality factor of the microscope is enhanced at low temperatures due to
the decrease of microwave losses in the resonator.

{\bf Model:} The inset of figure 1 shows a closer look at the tip to sample
interaction. This interaction is modelled as an effective capacitance C$_x$
in series with the losses in the conducting sample due to ohmic dissipation,
R$_x$. The complex load impedance presented to the microscope is Z$_x$ = R$%
_x $+(1/i$\omega $C$_x$).

Our quantitative understanding of the microscope is based on a transmission
line model developed earlier\cite{Gus,Steinhauer5}. In this model, the
frequency shift and Q of the microscope depend both on C$_x$ and R$_x$. In
the region of interest, an estimate (discussed below) for the capacitance at
a height of 1 nm is C$_x$ $\simeq $ 10 fF, giving a capacitive reactance
(Im[Z$_x$]) on the order of 2 k$\Omega $ at 7.5 GHz. For situations where C$%
_x$ is not too large (roughly C$_x$ values $\leq $ 10 fF), we can
approximate the model frequency shift as $\triangle $f = - b*C$_x$,
independent of R$_x,$ where b depends on microscope geometry. The model Q
depends on C$_x^2$ as Q = Q$_{\max }$ - d(R$_x$)*C$_x^2$, where Q$_{\max }$
is the microscope Q with no sample present. For increasing R$_x$ this slope
increases in magnitude, roughly as d(R$_x$) $\sim $ R$_x$. To summarize, in
this small capacitance limit, the frequency shift image can be regarded as a
capacitance image, and the Q image will contain contributions from both
capacitance C$_x$ and losses R$_x$. Similar results are obtained from a
lumped element model in which the resonator is treated as a parallel RLC
circuit.

In both models of the microscope we find in general that the minimum in Q
versus R$_x$ is always at the point where $\omega $C$_x$R$_x$ = 1. Hence the
sensitivity of our microscope to sample losses is determined in part by the
value of the probe-sample capacitance. Other observations about the
qualitative behavior of $\Delta $f and Q with sample properties have been
discussed at length in prior work\cite{Steinhauer4,Gus}.

To summarize, the frequency shift image should be a map of probe-sample
capacitance. Qualitatively, we expect C$_x$, to be large in a valley and
small near a peak on the sample surface, as shown in Fig. 2. The capacitance
between tip and sample can be calculated by assuming that the tip acts like
a metallic sphere above a metallic infinite plane\cite{Gao}. Naively, we
expect the spatial resolution for capacitance variations to be on the order
of the radius of the sphere.

{\bf Results: }We find that the value of capacitance between tip and sample
strongly depends on the geometry of the tip. We have used Pt-Ir alloy cut
tips, as well as Pt-Ir alloy etched tips and W etched tips. All three tips
have significantly different geometries. The W tip shows the largest $\Delta 
$f contrast as a function of height between tunneling and 2000 nm from the
surface (Fig.3). For this tip we have seen a frequency shift slope d($\Delta 
$f)/dz$\mid _{z\longrightarrow 0}$ contrast of roughly 0.3kHz/nm over a thin
gold film deposited on a mica substrate. The etched Pt-Ir tip has a smaller
contrast of 0.075kHz/nm, but still larger than a cut Pt-Ir tip, which has a
contrast of 0.025kHz/nm, all on the gold on mica thin film (all three tips
are compared in inset of Fig. 3). The largest frequency shift that we have
seen is 800 kHz between tunneling height and 500 nm, over an oxidized
Titanium sample with an etched W tip.

To quantitatively understand the frequency shift versus height data, we
calculate capacitance for a given sphere radius and height above the sample
starting from a typical tunneling height of 1 nm and extending to 2000 nm.
The values are fed into the transmission line model\cite{Gus,Steinhauer5},
which calculates the frequency shift. The inset of Figure 3 shows the fit
based on this model to the frequency shift versus height data for etched W
and Pt tips over the Gold/mica sample. The sphere radius was used as the
fitting parameter, and Fig. 3 shows that the W tip data fits well for a
sphere of radius 27 $\mu $m, while the etched Pt tip fits with a sphere of
10 $\mu $m. These values for tip radius fits are comparable to those seen by
other researchers\cite{Cho2,Gao} and give probe-sample capacitance on the
order of 1 - 10 fF at tunneling heights. We find however, that the Pt-Ir cut
tip is irregular and does not fit the sphere above the plane model.

Figure 4 shows simultaneously acquired images of STM topography and
microwave properties on a La$_{0.67}$Ca$_{0.33}$MnO$_3$ CMR thin film. The
STM topography clearly shows the granular structure of the film. The total
height variation is about 175\AA , and the smallest grain is about 285 \AA\ %
on each side. The simultaneously acquired frequency shift data shows all the
same granular features, with similar spatial resolution. Note that the
frequency shift is more negative for the region between the grains, and less
negative for regions near the top of the grains, as expected from Fig. 2.

Surprisingly, the $\Delta $f and Q image spatial resolution is just as good
as STM topography. The contrast in $\Delta $f and Q come from the
topography-following mode, where STM feedback is maintaining a constant
tunnel current. As the tip goes into a valley on the surface, the microwave
microscope will see an increase in capacitance between the tip and sample
(Fig. 2) which will produce a more negative frequency shift. A stronger drop
in Q is also seen due to the increase in C$_x$ and possibly also due to a
change in R$_x$, as proposed\cite{Amlan} for CMR thin films.

Figure 5 shows the data on the top of one grain of the La$_{0.67}$Ca$_{0.33}$%
MnO$_3$ film, shown in Fig. 4. This image is 492\AA\ on each side and the
overall STM topography is 96\AA . The simultaneously acquired frequency
shift image ranges from -105kHz to -110kHz and the Q image ranges from 348
to 357. The dark lines in the topography image are narrow dips about 8 to 10
unit cells deep. The Q and frequency shift images clearly show these dips as
well. Figure 6 shows the line cut through the largest dip in Fig. 5, where
the topography shows that this feature is 55\AA\ deep, from the top of the
grain. The frequency shift change is about 2 kHz and Q drops from 356.5 to
348.5 over this feature. One can regard Fig. 5 as a relatively flat region
of the sample where the microscope shows a baseline frequency shift, $\Delta 
$f $\sim $ -105 kHz. When the tip moves into the 5.5 nm deep valley (Fig.
6), under STM constant-current mode, the microwave microscope shows an
additional drop of $\sim $ 2 kHz in frequency shift, giving a slope, d($%
\Delta $f)/dz $\sim $ 0.3 kHz/nm, consistent with the results shown in Fig.3
for the frequency shift contrast near the surface.

This analysis suggests, that the STM constant-current mode is needed to see
such sharp contrast in frequency shift, due to changes in capacitance.
Hence, the spatial resolution will be dictated by the STM constant-current
mode rather than the radius of curvature of the tip, as expected from the
simple model of microwave microscope. Comparing the three line cuts through
the feature (in Fig. 6), we clearly see that the spatial resolution for
capacitance variations of the microwave microscope is comparable to STM, and
is no worse than 25\AA !

Noise ultimately limits our sensitivity to capacitance variations. There is
noise in both the STM positioning system and the microwave microscope. At
room temperature the estimated position noise of the z piezo is 0.35\AA\ and
the position noise in x and y direction is 1.2\AA . This translates to
0.0026 kHz of noise in $\Delta $f and 0.0083 in Q for a typical Pt-etch tip,
so we conclude that the contribution to noise from the positioning system is
negligible. We find that for the microwave microscope, sitting far away from
the sample the jitter seen in the frequency shift signal is about 0.5kHz,
and the variation in Q is about 0.1 out of 383. The lock-in time constant
was 1 ms for these experiments.

{\bf Conclusions:} We have demonstrated a novel microwave frequency scanning
capacitance microscope with spatial resolution of no worse than 25\AA . The
microwave contrast depends strongly on the tip-sample capacitance. This
capacitance depends strongly on the geometry of the tip. This scanning
capacitance microwave microscope can serve as a high resolution platform for
doing other kinds of measurements, such as local loss and local nonlinear
properties\cite{SCLee} of interesting materials.

This work has been supported by an NSF SBIR-II subcontract from Neocera,
Inc. under NSF DMI-0078486, an NSF Instrumentation for Materials Research
grant DMR-9802756, the University of Maryland/Rutgers NSF-MRSEC shared
experimental facility under grant number\ DMR-00-80008, the Maryland
Industrial Partnerships Program 990517-7709, and by the Maryland Center for
Superconductivity Research. We acknowledge Amlan Biswas for providing us
with the La$_{0.67}$Ca$_{0.33}$MnO$_3$ film, as well as Greg Ruchti for
finite-element electromagnetic modeling of the microscope.

\begin{figure}[tbp]
\caption{Schematic diagram of the STM-assisted scanning near field microwave
microscope. A model of the probe-sample interaction is shown in the inset.}
\label{schematic}
\end{figure}

\begin{figure}[tbp]
\caption{Schematic of a)tip above flat region of the sample, b)tip in a
valley and c)tip above a grain. The relative capacitance values between the
tip and the sample are C$_{peak}$ $<$ C$_{flat}$ $<$ C$_{valley}$.}
\label{image1}
\end{figure}

\begin{figure}[tbp]
\caption{Comparison of frequency shift vs. distance from tunneling height
data for Pt-Ir etched (open circles) and W etched tips (open triangles) to
the sphere above the plane model (solid lines). Experiments were performed
at 7.37GHz. Inset plots the frequency shift versus log z to show the
logarithmic distance scaling from the sphere above the plane capacitance model. The
data are normalized to $\triangle $f (2000 nm) = 0 kHz. The inset also shows
data for Pt-Ir cut tip (hollow squares), performed at 7.25 GHz. The sample is a thin 
Gold film on mica substrate.}
\label{image2}
\end{figure}

\begin{figure}[tbp]
\caption{Simultaneously acquired a) Topography, b) Quality Factor and c)
frequency shift image of La$_{0.67}$Ca$_{0.33}$MnO$_3$, thin film which is
1000 \AA\ thick on LaAlO$_3$ substrate. The images are 6000 \AA\ on each
side. Data is taken att 272K with a Pt-Ir etch tip at 7.67 GHz.}
\label{image3}
\end{figure}

\begin{figure}[tbp]
\caption{Images of the top of one grain of a La$_{0.67}$Ca$_{0.33}$MnO$_3$
a)STM topography b)Quality factor and c)frequency shift images. Image size
is 492 \AA\ on each side. The horizontal light and dark wide stripes on the
topography image are due to temperature drift. Data is taken at 240K with a 
Pt-Ir etch tip at 7.67 GHz. The horizontal line cut in a) is shown in Fig. 6.}
\label{image4}
\end{figure}

\begin{figure}[tbp]
\caption{Line cut of the data shown in Figure 5. It demonstrates a 25 \AA\
spatial resolution for capacitance variations in the microwave response of
the sample.}
\label{image5}
\end{figure}

\end{document}